**Tunable strong coupling of two adjacent optical *λ*/2 Fabry-Pérot microresonators**

Achim Junginger[1], Frank Wackenhut[1,*], Alexander Stuhl[1,2], Felix Blendinger[3], Marc Brecht[2] and Alfred J. Meixner[1,*]

[1]Institute of Physical and Theoretical Chemistry, Eberhard Karls University, 72076 Tübingen, Germany

[2]Process Analysis and Technology, Reutlingen University, 72762 Reutlingen, Germany

[3]Faculty for Mechanical and Medical Engineering, Furtwangen University, 78054 Villingen-Schwenningen, Germany

ABSTRACT

Optical half-wave microresonators enable to control the optical mode density around a quantum system and thus to modify the temporal emission properties. If the coupling rate exceeds the damping rate, strong coupling between a microresonator and a quantum system can be achieved, leading to a coherent energy exchange and the creation of new hybrid modes. Here, we investigate strong coupling between two adjacent *λ*/2 Fabry-Pérot microresonators, where the resonance of one microresonator can be actively tuned across the resonance of the other microresonator. The transmission spectra of the coupled microresonators show a clear anticrossing behavior, which proves that the two cavity modes are strongly coupled. Additionally, we can vary the coupling rate by changing the resonator geometry and thereby investigate the basic principles of strong coupling with a well-defined model system. Finally, we will show that such a coupled system can theoretically be modelled by coupled damped harmonic oscillators.

Optical $\lambda/2$ microresonators are structures that confine light to volumes with dimensions on the order of a wavelength and enable to control and study light-matter interaction. The interaction between a quantum system and an optical field confined in a microresonator can be divided into the weak and strong coupling regime. In the weak coupling regime, the respective decay rates are larger than the coupling rate between the quantum system and the microresonator. In this case, the spontaneous emission rate of the quantum system is altered with respect to the free space, a phenomenon known as Purcell effect [1]. To reach the strong coupling regime, the coupling strength between the optical field in the resonator and the quantum system must be considerably larger than their respective decay rates. This leads to new hybrid polaritonic states [2], which have an energy difference proportional to the coupling strength. The spectral signature is a splitting of the absorption or transmission spectrum into two polaritonic modes, referred to as Rabi splitting [3]. When the cavity resonance is tuned over the eigenfrequency of the quantum system, anticrossing is observed in the dispersive behavior of the polaritonic modes [4]. The first observation of strong coupling between electromagnetic fields and a quantum system has been shown in the form of interaction between Rydberg atoms and a high Q microwave cavity at cryogenic temperatures [5]. Since then, many different optical experiments showing strong light matter coupling have been accomplished using metal or dielectric cavities [4,6-11], photonic crystals [12], micropillars [13] or microdisks [14] that couple with quantum dots [12-14], organic semiconductors [15] or J-aggregates [16]. Strong coupling has been shown for molecular systems from ensembles down to single molecules that couple to cavity fields, as well as to plasmonic modes [17-19] with sub wavelength dimensions. Today, strong coupling with plasmonic modes at ambient conditions has been shown even for single molecules [20]. Recently, strong coupling has been used to influence chemical reactions, e.g. by strong coupling of molecular vibrations to an infrared cavity, the chemical reaction rate involving this particular vibration can be altered [21]. More examples can be found in recent review articles [4,22], including strong coupling in Fabry-Pérot type bare metal microresonators [6]. Apart from light matter interaction, strong coupling is also an important effect in cavity optomechanics [23], in coupled microdisks [24,25] and classical physics as in coupled oscillators [26].

In former studies, we have used Fabry-Pérot microresonators consisting of evaporated silver mirrors separated by half an optical wavelength to influence the emission rates and fluorescence spectra of single perylene-type molecules [27-29] or nanoparticles [30]. We also used microresonators to control the energy transfer of single FRET pairs [31-33] or larger systems like photosystem I [34]. These effects can be described by the Purcell effect and are in the weak coupling regime. In another experiment, we have also shown strong coupling of individual plasmonic gold nanoparticles with optical modes of a low Q microresonator at ambient conditions [35]. Here, we will study the mode-coupling between two adjacent optical microresonators consisting of three mirrors with a mirror spacing of half a wavelength, suitable for resonances in the visible spectral region. We will show that such a microresonator is ideal to study strong coupling effects since the coupling strength can be tailored by changing the properties of the central mirror.

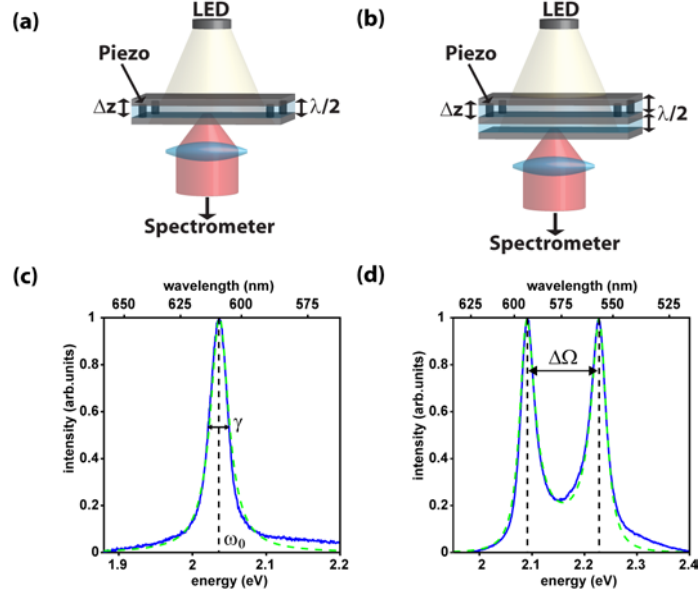

FIG. 1. Schematic drawing of a single tunable microresonator consisting of two parallel silver mirrors serving as a reference (a) and two coupled resonators, which are separated by a partially transmitting silver mirror (b). The resonators are illuminated with a white light LED from the top. Piezo actuators allow to tune the optical path length Δz with high precision within the λ/2 region of the visible spectral range. In the coupled resonator, shown in (b), the mirror spacing of the lower resonator is fixed and actuators allow to tune the optical path length Δz of the upper resonator. (c) Transmission spectrum (blue line) of the single microresonator, which can be fitted by a Lorentzian line shape (green dashed line) of a harmonic oscillator with a full width at half maximum of $\gamma$ = 30 meV and with a resonance at $\omega_0$ = 609 nm fitted to the data. (d) Transmission spectrum of the coupled microresonators (blue line) with two transmission maxima fitted by the spectral line shape (green dashed line) of two coupled harmonic oscillators that have identical eigenfrequencies $\omega_0$, separated by the Rabi splitting $\Delta\Omega$.

As a reference, we consider first a single microresonator (Fig. 1 (a)) consisting of two silver mirrors separated by half a wavelength in the visible spectral region [31], with a resonance that can be precisely tuned by changing the mirror separation Δz with piezo actuators. The mirrors are fabricated from microscopy cover slides by electron beam evaporation of a 50 nm thick silver layer, followed by a 10 nm gold layer and a protection layer of 10 nm $SiO_2$. The final resonator structure is assembled in a home-built holder with piezo actuators (KC1-PZ/M, Thorlabs) and immersion oil (Immersol 518F, Zeiss) between the two mirrors. The piezo actuators allow to precisely tune the cavity resonance by moving the upper mirror by a well-defined distance Δz in steps of down to 5 nm. The second configuration consists of two coupled resonators schematically shown in Fig. 1 (b) and is an extension of the single microresonator described above. The top mirror is identical, but the lower mirror is replaced by a microresonator with a fixed optical path length. It consists of a 50 nm thick silver layer on top of the lower cover slip followed by a transparent $SiO_2$ spacer layer of about 150 nm thickness which is covered with a silver layer of variable thickness. This silver layer forms a mirror that is shared by both resonators. Additionally, a 10 nm thick gold and a 10 nm $SiO_2$ layer are used to protect the central mirror against the immersion oil in the upper resonator. The central silver layer has a thickness of 14 nm, 24 nm or 38 nm giving reflectivities of 66 %, 85 % and 95 %, respectively, resulting in different coupling constants κ between the upper and lower resonator. Transmission spectra are recorded by illuminating the microresonator through the upper mirror with a white light LED operating under continuous wave conditions. The transmitted light is collected from below with a home built confocal microscope equipped with an oil immersion objective lens (Zeiss alpha Plan-Apochromat 63x/1,46 oil) and a spectrometer equipped with a thermoelectrically cooled CCD-Detector. Figure 1 (c) and (d) shows the respective experimental (blue line) and simulated (green dashed line) transmission spectra of the single and coupled microresonator, respectively. For a single microresonator only one Lorentzian shaped transmission peak is observed, while the coupled system shows two transmission peaks separated by the Rabi splitting $\Delta\Omega$.

Since the transmission spectrum can be fitted, for stationary conditions, by a Lorentzian line shape function we may describe the autocorrelation function of the transmitted signal by a damped harmonic oscillator with an amplitude given by:

$$x(t) = e^{-\frac{\gamma t}{2}}[\cos(\omega_d t)], \omega_d = \sqrt{\omega^2 - \left(\frac{\gamma}{2}\right)^2} \tag{1}$$

with the resonance frequency $\omega_d$ and the damping constant $\gamma$. Figure 2 (a) displays the analytical solution $x(t)$ in red and the Fourier transform of the time dependent amplitude $x(t)$ is shown by the red dashed line in Fig 2 (b). The blue lines in Fig. 2(a)/(b) are the respective numerical solutions for a single damped harmonic oscillator. This approach gives a Lorentzian line shape (green line) which is in perfect agreement with the experimental data (blue line) shown in Fig. 1 (c). In the following, we model the autocorrelation function of the coupled system by two coupled damped harmonic oscillators for which the respective power spectral density can again be calculated by Fourier transformation. The equation of motion for the amplitude of the two coupled harmonic oscillators are described by two coupled differential equations, which can be written as:

$$\begin{aligned} \ddot{x}_1(t) + \gamma_1 \dot{x}_1(t) + \omega_1^2 x_1(t) + \kappa x_2(t) = 0 \\ \ddot{x}_2(t) + \gamma_2 \dot{x}_2(t) + \omega_2^2 x_2(t) + \kappa x_1(t) = 0 \end{aligned} \tag{2}$$

with the damping constants $\gamma_1, \gamma_2$, the resonance frequencies $\omega_1, \omega_2$ of the two individual oscillators and the respective coupling constant $\kappa$.

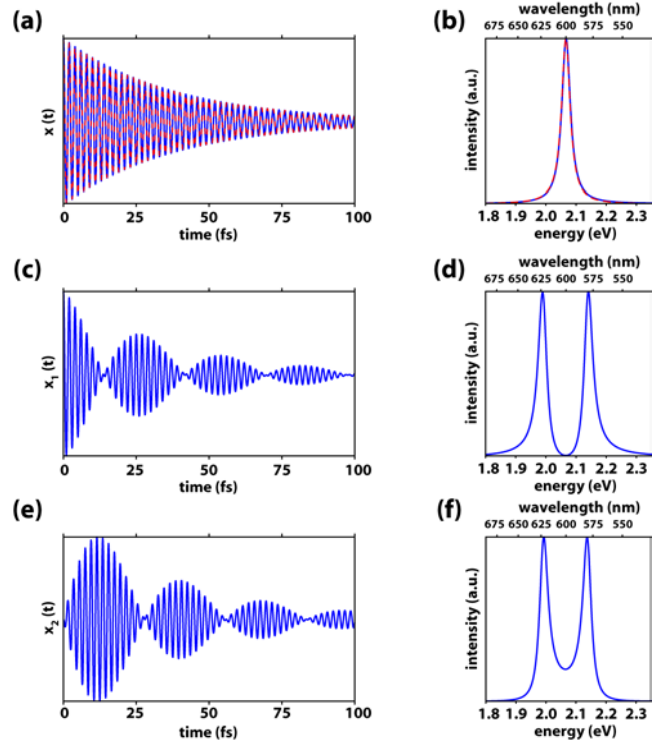

FIG. 2. (a) Analytical (red, Eq. (1)) and numerically calculated (blue, from Eq. (1)) decay of the amplitude x(t) of a damped harmonic oscillator with a resonance frequency ω corresponding to a wavelength of 600 nm and a damping constant γ=34.5 meV. The blue spectrum shown in (b) is the Fourier transform of x(t) shown in (a) and has a single peak at ω (expressed in eV) with a full width at half maximum of γ. The red curve is a Lorentzian shaped analytical solution and validates the procedure. (c) and (e) illustrate the temporal response of two coupled oscillators with the same resonance frequency and damping constant, where a beating pattern can be observed due to the energy exchange between the oscillators. The spectra in (d) and (f) are the respective Fourier transformations of (c) and (e) showing two intense maxima caused by strong coupling of the oscillators.

By solving these equations numerically, we obtain $x_1(t)$ and $x_2(t)$, which are shown in Fig. 2 (c) and (e), respectively. The coupled system is illuminated from the top, therefore we set the starting amplitude for the first oscillator to $x_1(0) = 1$, while it is zero for the second one $x_2(0) = 0$. In this case the second oscillator $x_2(t)$ is exclusively excited via the coupling to $x_1(t)$. After the excitation of $x_1(t)$ the energy is transferred to $x_2(t)$ and since the transfer is allowed in both directions the energy is transferred back to $x_1(t)$. Due to this coherent energy exchange, caused by the coupling of the two harmonic oscillators, we can observe a beating pattern in the temporal response $x_1(t)$ and $x_2(t)$. Again, the power spectral density of the coupled resonators is proportional to the Fourier transforms of $x_1(t)$ and $x_2(t)$ and are represented for the special case of $\omega_1 = \omega_2$ in Fig. 2 (d) and (f) showing two resonator modes in the spectrum of $X_1(\omega)$ and $X_2(\omega)$, which are separated by the Rabi splitting $\Delta\Omega$.

We can now use these equations and set the resonance wavelengths and damping constants to fit the experimental data shown in Fig. 1 and find a perfect agreement for both the single and coupled system. An exemplary spectrum of the coupled microresonator is shown in blue in Fig. 1 (d) together with a simulation based on two coupled harmonic oscillators with $\lambda_1 = \lambda_2$ =573.4 nm, $\gamma_1 = \gamma_2$ =34.5 meV and κ=0.465 eV. By comparing the experimental spectrum to $x_1(\omega)$ and $x_2(\omega)$ we find a perfect agreement with $x_2(\omega)$, which can be explained since we detect from below and only light fulfilling the resonance condition of the coupled microresonators can reach the detector.

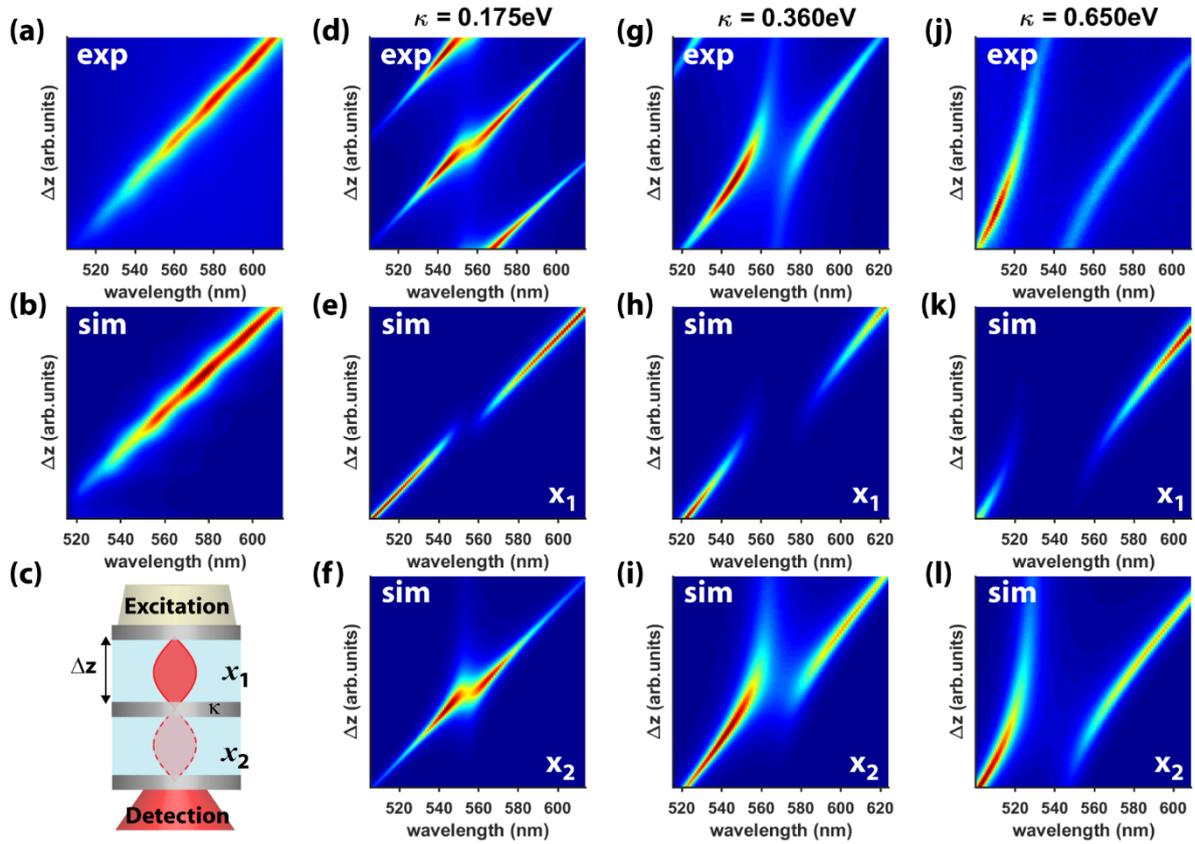

FIG. 3. (a) Experimental transmission spectra for an uncoupled microresonator where the transmission maximum is tuned by moving the upper mirror by a distance Δz. (b) The corresponding simulations to (a) for a single microresonator. (c) Schematic drawing of the coupled microresonator, where $x_1$ and $x_2$ describe the two coupled resonator modes. (d) Transmission spectra for a coupled microresonator as depicted in (c) and the respective simulations for $x_1$ and $x_2$ are shown in (e) and (f). In this particular case the central mirror has a thickness of 38 nm resulting in a coupling constant of κ = 0.175 eV and an anticrossing dispersion can be observed. (g) and (i) Similar experimental and simulated results for a thickness of the central mirror of 24 nm and the Rabi splitting is increased compared to (d) and (f). (j)-(l) The thickness of the central mirror is even further decreased to 14 nm leading to a larger Rabi splitting.

In order to further prove that the splitting observed in Fig. 1 (d) is indeed caused by strong coupling we have investigated the dispersive behavior of the microresonator modes. This can be achieved by tuning the resonance of the upper resonator across the resonance of the lower one by moving the upper mirror. The results of this experiment are illustrated in Fig. 3. For comparison, Fig. 3 (a) shows experimental data for a single microresonator where the resonance at the focal position of the microscope is tuned from 505 nm to 611 nm. The intensity modulations seen in the experimental spectra is caused by the emission profile of the white light LED and is also considered in the simulations. Figure 3 (b) displays the corresponding simulation of the single microresonator where the resonances are adjusted to match the experimental data shown in Fig. 3 (a). We find a perfect agreement between the simulated and experimental dispersion, which shows again that harmonic oscillators can be utilized to model such microresonator systems. Figure 3 (c) presents a schematic representation of the resonator geometry where the resonance of the upper resonator can be tuned by moving the topmost mirror by a defined distance Δz. The excitation of the coupled system is achieved by illumination with a white light LED through the topmost mirror, while the transmission signal is collected with an objective lens from below. Figure 3 (d) shows experimental transmission data of such a coupled microresonator system with a 38 nm thick central mirror. Here, the resonance of the upper microresonator $\omega_1$ is tuned across the fixed resonance $\omega_2$ of the lower one and an anticrossing behavior can be observed when $\omega_1$ is close to $\omega_2$. The Rabi splitting of this coupled resonator system is measured to be ΔΩ = 7.9 nm (31.9 meV). Note, that a mode of a lower and a higher order can also be seen in the experimental spectra for low and high Δz values, these modes have not been considered in the simulations. We can model such a microresonator geometry with two coupled harmonic oscillators following Eq. (3), where the upper resonator is described by $x_1(t)$ with a tunable frequency $\omega_1$ and the lower resonator is modeled by $x_2(t)$ with a fixed frequency $\omega_2$. The respective simulations are shown in Fig. 3 (e) and 3 (f), where the spectral response of $x_1$ is presented in (e) and $x_2$ in (f). From the fit of the theoretical model to the experimental data we obtain the damping constants and the coupling constants for $\gamma_1 = 11$ meV, $\gamma_2 = 50$ meV and κ = 175 meV. Again, the response of $x_2$ perfectly matches the experimental data, which can be explained by the fact that we excite the coupled resonator from the top and collect the transmission signal from below and only light from the lower resonator will reach the detector. We can separate the excitation of the lower resonator into direct excitation and excitation via coupling to the upper resonator. A direct excitation can be caused by leakage of the white light through the upper resonator structure due to the finite reflectivity of the upper two mirrors. However, this portion is in the range of a few percent (depending on the actual mirror thicknesses) of the incoming white light intensity since it is reflected by the two topmost mirrors. Furthermore, most wavelengths of the white light spectrum do not reach the lower resonator since they do not fulfill the resonance condition of the upper resonator. Therefore, in contrast to the single microresonator, we did not consider the spectral profile of the white light LED to reproduce the experimental data because the white light spectrum is prefiltered by the upper resonator. This creates the situation that there is only a weak direct excitation of the lower resonator, which can be taken into account in the simulation by modifying the starting conditions for the two oscillators. We set $x_1(0) = 1$ for the first oscillator, since it is directly excited by the white light LED, and the weak direct excitation of $x_2$ is taken into account by setting $x_2(0) = 0.05$. However, this small change of the starting conditions, due to the weak direct excitation of $x_2$, results in an intensity difference between the two coupled resonator modes, which is also observed in the experimental data. The second excitation pathway reflects the coherent energy exchange between the upper and lower resonator due to strong coupling, which leads to the observed anticrossing behavior of the two resonator modes. We find an excellent agreement between the experimental and the simulated data for these starting parameters leading to the conclusion that the lower resonator is mainly (95%) excited by the coupling to the upper resonator and that the energy is coherently exchanged between the two resonators. In Fig. 3 (g) experimental data is shown where the thickness of the central mirror is reduced from 38 nm to 24 nm, which increases the coupling between the two resonators and results in an increase of the Rabi splitting between the two resonator modes from 7.9 nm (31.9 meV) to 25.7 nm (99.6 meV). The corresponding simulations for $x_1$ and $x_2$ are shown in Fig. 3 (h) and (i) and the parameters used for the simulations are $\gamma_1 = 13$ meV, $\gamma_2 = 65$ meV and κ = 360 meV, which shows that the reduction of the central mirror thickness leads to an increase of the coupling constant from κ = 175 meV to κ = 360 meV. Again, we find the best match between the response of $x_2$ and the experimental data. In this case the thickness of the central mirror is reduced and the portion of direct excitation of $x_2$ is larger ($x_1(0) = 1$, $x_2(0) = 0.1$), which results in a stronger intensity difference between the two coupled modes. This effect is even more pronounced when the central mirror thickness is further reduced to 14 nm, which is experimentally shown in Fig. 3 (j) and the respective simulations are presented in Fig. 3 (k) and Fig. 3 (l). In this case the coupling strength between the two resonators is even larger and the Rabi splitting increases to 33.3 nm (146.1 meV). The parameters used in the simulation are $\gamma_1 = 13$ meV, $\gamma_2 = 55$ meV and $\kappa = 650$ meV and again we find the best match between $x_2$ and the experimental data.

These results show that we have created different microresonator structures where the coupling constant can be tuned by a large amount from 175 meV to 650 meV making them ideal to study the fundamental principles of strong coupling. Furthermore, a damped harmonic oscillator approach is sufficient to model such coupled microresonator structures and extract important parameters, i.e. the damping and coupling constants.

In summary, we prepared coupled λ/2 optical resonators which show strong coupling between the respective modes. The coupling strength can be adjusted by varying the thickness of the central silver mirror. Furthermore, we have shown that we can use the equations of motion of coupled damped harmonic oscillators to theoretically model such a strongly coupled system. For stationary conditions the white light transmission signal can be modelled by the Fourier transform of the time domain signal of the second microresonator which is strongly coupled to the first resonator. Such a system can be used to manipulate the mode structure in the fixed microresonator without changing its geometry but by tuning the upper resonator. This may lead to exciting new applications with tunable subwavelength structures in the rapidly growing field of nanoswitches and optoelectronics.

ACKNOWLEDGMENTS

The authors thank Michael Metzger and Alexander Konrad for their contribution in the initial phase of this project. This work was financially supported by the Federal Ministry of Science, Research and Arts of Baden-Württemberg (Kooperatives Promotionskolleg „IPMB") and the DFG (ME 1600/13-3).

* Corresponding author. frank.wackenhut@uni-tuebingen.de, alfred.meixner@uni-tuebingen.de